# Optimal Power Flow in Hybrid AC and Multi-terminal HVDC Networks with Offshore Wind Farm Integration Based on Semidefinite Programming

Yuhao Zhou
Long Zhao
Wei-Jen Lee
Dept. of Electrical Engineering
ESRC in the University of Texas at Arlington
Arlington, TX, U.S.A.
yuhao.zhou@mavs.uta.edu

Zhenyuan Zhang
Peng Wang
School of Mechanical and Electrical Engineering
University of Electronic Science and Technology of China
Chengdu, China

*Abstract*—Multi-terminal high voltage direct current (MT-HVDC) technology is a promising technology for the offshore wind farm integration, which requires the new control and operation scheme. Therefore, the optimal power flow problem for this system is important to achieve the optimal economic operation. In this paper, convex relaxation model based on semidefinite programming for the MT-HVDC system considering DC/DC converters is proposed to solve the optimal power flow problem. A hybrid AC and MT-HVDC system for offshore wind farm integration is used for the test. The simulation results validate the effectiveness of the proposed model and guarantee that the global optimum solution is achieved.

*Index Terms*-- Convex relaxation, multi-terminal high voltage direct current (MT-HVDC), optimal power flow, offshore wind farm, semidefinite programming (SDP).

## I. NOMENCLATURE

| | |
|---|---|
| $AC, DC$ | Index for AC and DC system |
| $\mathcal{J}$ | Set for generator buses in AC system |
| $\mathcal{N}$ | Set for buses in the system |
| $\mathcal{L}$ | Set for transmission lines in the system |
| $\mathcal{D}$ | Set for buses connected with DC/DC converter |
| $\mathcal{O}$ | Set for AC buses connected with AC/DC converters |
| $Y$ | Admittance matrix in the system |
| $n$ | Bus number of the system |
| $e_n$ | $n^{th}$ basis vector in $\mathbf{R^n}$ |
| $V$ | Voltage vectors of buses in the system |
| $P_{Gk}$ | Generator injected power for bus $k$ |
| $P_{Dk}$ | Load injected power for bus $k$ |
| $P_{Gk}^{min}$ | Minimum active power output for generator $k$ |
| $P_{Gk}^{max}$ | Maximum active power output for generator $k$ |
| $Q_{Gk}^{min}$ | Minimum reactive power output for generator $k$ |
| $Q_{Gk}^{max}$ | Maximum reactive power output for generator $k$ |
| $P_k^{min}$ | Minimum injected active power for bus $k$ |
| $P_k^{max}$ | Maximum injected active power for bus $k$ |
| $Q_k^{min}$ | Minimum injected reactive power for bus $k$ |
| $Q_k^{max}$ | Maximum injected reactive power for bus $k$ |
| $V_k^{min}$ | Minimum voltage requirement for bus $k$ |
| $V_k^{max}$ | Maximum voltage requirement for bus $k$ |
| $S_{lm}^{max}$ | Maximum apparent power for the line between bus $l$ and $m$ |
| $P_{lm}^{max}$ | Maximum active power for the line between bus $l$ and $m$ |
| $c_{k2}, c_{k1}, c_{k0}$ | Quadratic cost coefficients for generator $k$ |
| $g_{ft}$ | The conductance of the HVDC line between bus f and t |
| $\eta_k$ | Efficiency for AC/DC converter $k$ |
| $s_k$ | Converter loss at bus $k$ caused by DC/DC converter in MT-HVDC system |
| $\delta_k, \beta_k, \gamma_k$ | Converter loss factors of $k^{th}$ DC/DC converter in MT-HVDC system |
| $q_{lm}$ | Power in DC/DC converter between bus $l, m$ |
| $S$ | Converter loss matrix in HVDC system |
| $Q$ | Power flow vector for DC/DC converters |
| $A_d$ | Connection matrix for DC/DC converters |
| $P_{wind\_k}$ | Injected power in bus $k$ from wind farm |
| $\lambda_1, \lambda_2$ | Largest and second largest eigenvalue of matrix |
| $E_1, E_2$ | The corresponding eigenvector of $\lambda_1$ and $\lambda_2$ |

## II. INTRODUCTION

Wind energy has become an important electric power source nowadays due to its cost-effectiveness and environmental friendliness [1]. In addition to the onshore wind farms, offshore wind farms also draw a lot of attention because of better wind quality and limited inland sites to build new wind farms [2]. High voltage direct current (HVDC) technology is one of the most favorable options for the integration of offshore wind farms to deliver the power for inland customers [3]. Multi-terminal HVDC grids are emerging as a very promising concept to support the mesh DC networks such as SuperGrid [4] proposed in Europe, and therefore draws a lot of interests from industry and academia. In China, it ranked the 3rd in terms of total installed capacity of offshore wind turbines, and a new 1160 MW capacity of offshore wind farms are installed in 2017, an increase of 97% over the previous year [5]. Compared with point to point HVDC transmission system, MT-HVDC are based on highly controllable devices such as voltage source converter (VSC) based terminals, which not only could transmit the power but also supporting AC grids to ensure a security and stable operation. This technology has the capability to transform the traditional inverter from grid following to grid forming. However, the MT-HVDC system is more difficult to control. In addition, highly meshed MT-HVDC systems require DC/DC stations to interconnect HVDC systems with different nominal voltage or different configuration such that it could control the power flow on a specific HVDC line, which makes them equally important in MT-HVDC grids [6]. Therefore, the optimal power flow (OPF) problem for the new type MT-HVDC grids integrated with offshore wind farms is very critical for future power system operation and control, which could achieve the economic operation while maintain the limits.

Non-convex OPF problem has been extensively studied recently to search the global optimum solution. In [7]-[9], it is proved that the global optimum point could be achieved if the duality gap is zero by applying semidefinite programming (SDP) relaxation. In [10], a SDP based method is proposed to solve the OPF problem for hybrid systems with DC micro-grids. In [11], an improved approach to solve the security constraint OPF for MT-HVDC system is proposed, while the method is based on local optimal approach. In [6], a SDP based method is proposed to solve the OPF problem only for MT-HVDC system, while the interconnection with AC system and offshore wind farm is not studied.

In this paper, the global optimum solution of the OPF problem is discussed for the hybrid AC and MT-HVDC system considering offshore wind farm integration. The non-convex OPF problem is transformed to convex problem by applying SDP convex relaxation. The simulation results verify the effectiveness of the proposed method.

The organization of this paper is discussed as follows. Section III describes the non-convex form OPF for the hybrid AC and MT-HVDC system, section IV transforms the non-convex problem into convex problem by applying SDP relaxation. And a case study is shown in section V to prove the effectiveness of the proposed method.

### III. MODELS FOR HYBRID AC AND MT-HVDC SYSTEM WITH OFFSHORE WIND FARM INTEGRATION

The Hybrid AC and MT-HVDC systems consists four major parts: the AC system, MT-HVDC system with DC/DC converters, AC/DC converters and the offshore wind farm. The standard power flow equations are used to formulate the OPF problem. In this section, these four systems are modeled as follows.

#### A. AC Systems

Consider an AC system with a set of buses $N_{AC} = \{1, 2, ..., n_{AC}\}$, a set of transmission lines $L_{AC}$, and the generator buses are $g_{AC} \subseteq N_{AC}$, then the classic OPF problem could be formulated as follows [7]-[9]:

$$\begin{aligned}
& \text{minimize} && \sum_{k \in \mathcal{G}_{AC}} f_k(P_{Gk\_AC}) \\
& \text{subject to} \\
& P_{Gk\_AC}^{\min} \leq P_{Gk\_AC} \leq P_{Gk\_AC}^{\max}, \forall k \in \mathcal{G}_{AC} \\
& Q_{Gk\_AC}^{\min} \leq Q_{Gk\_AC} \leq Q_{Gk\_AC}^{\max}, \forall k \in \mathcal{G}_{AC} \\
& V_{k\_AC}^{\min} \leq |V_{k\_AC}| \leq V_{k\_AC}^{\max}, \forall k \in \mathcal{N}_{AC} \\
& \theta_{k\_AC}^{\min} \leq |\angle V_{k\_AC}| \leq \theta_{k\_AC}^{\max}, \forall k \in \mathcal{N}_{AC} \\
& |S_{lm\_AC}| \leq S_{lm\_AC}^{\max}, \forall (l,m) \in \mathcal{L}_{AC} \\
& P_{Gk\_AC} - P_{Dk\_AC} = \sum_{l \in \mathcal{N}_{AC}(k)} \text{Re}\{V_{k\_AC}(V_{k\_AC}^* - V_{l\_AC}^*) y_{kl}^*\} \\
& Q_{Gk\_AC} - Q_{Dk\_AC} = \sum_{l \in \mathcal{N}_{AC}(k)} \text{Im}\{V_{k\_AC}(V_{k\_AC}^* - V_{l\_AC}^*) y_{kl}^*\}
\end{aligned} \quad (1)$$

where $y_{kl}$ is the admittance between bus $k$ and $l$. The objective of this problem is to minimize the costs of generators, and $f_k(P_{Gk\_AC})$ is the quadratic cost function for generator $k \in \mathcal{G}_{AC}$ shown as follows:

$$f_k(P_{Gk\_AC}) = c_{k2} P_{Gk\_AC}^2 + c_{k1} P_{Gk\_AC} + c_{k0}, \forall k \in \mathcal{G}_{AC} \quad (2)$$

Because of the equality power flow equations in (1), this problem is not convex.

#### B. MT-HVDC System with DC/DC converter

Although there are many different topologies for the MT-HVDC system, the AC/DC terminals are modeled as VSC type in this paper. Because of the AC/DC terminals integrate AC grids and HVDC grids, AC/DC converters control the power injections or consumptions at that specific bus and can be treated as nodal devices. However, the power flow among the HVDC lines can't be controlled by AC/DC terminals alone. DC/DC converters is expected to be included in the MT-HVDC systems to enhance controllability [6]. Therefore, this new control variable must be considered in the power flow formulation.

There are several different control strategies for MT-HVDC system including master-slave control, margin voltage control, priority control, ratio control and droop control [2]. This paper adopts master-salve control scheme for AC/DC terminals. Therefore, one DC bus can be modeled as a slack node. An example of MT-HVDC system with one DC/DC converter based on a reduced version of the CIGRE B4 bipolar DC grid [6] are shown in Figure 1. In this system, there are 8 buses, 10 transmission lines and six AC/DC converters with one DC/DC converter, where one AC/DC converter connects with one offshore wind farm.

Firstly, the model of AC/DC converter is established as two generators connected with the corresponding AC bus and DC bus. An example is shown in Figure 2. The power balance of AC/DC converters needs to be considered as follows, where the power loss in converter is modeled by its efficiency.

$$\eta_i \times P_{ig\_AC} + P_{ig\_DC} = 0, \forall i \in \mathcal{O} \quad (3)$$

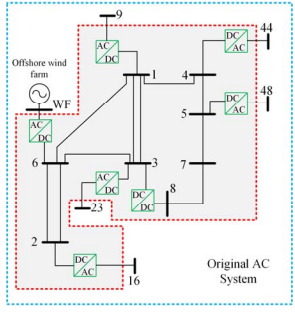

Figure 1. MT-HVDC system with one DC/DC converter

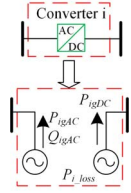

Figure 2. Equivalent model for AC/DC converter in MT-HVDC system

As for the bus $k \in \mathcal{O}$, there's another constraint due to the converter's capacity limit, which is shown in (4).

$$P_k^2 + Q_k^2 \leq S_{k\_conv}^2, \forall k \in \mathcal{O} \quad (4)$$

As for DC/DC converters, similar to [6], the losses are modeled as quadratic form related to the average transferred power, where the assumptions and power loss factors are based on [12]. Consider the DC/DC converter connected with node k and m, which is shown in Figure 3, the losses are modeled as two loads, which represent the half loss of that converter. The power balance is shown in equation (4).

$$q_{km} = -q_{mk}$$
$$s_k = \frac{1}{2}(\delta_k + \beta_k \|q_{km}\|_2 + \gamma_k \|q_{km}\|_2^2) \quad (5)$$

Because the power flow is calculated in steady state, the offshore wind farm in Figure 1 is modeled as one equivalent generator [1], and treated as PQ bus type. Then the power flow for MT-HVDC system could be formulated as follows:

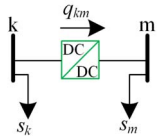

Figure 3. Losses model for DC/DC converter in MT-HVDC system

$$\begin{aligned}
&V_{1\_DC} = V_{master} \\
&P_{k\_DC} - s_k = \sum_{l \in \mathcal{N}_{DC}(k)} a_{kl} q_{kl} + \sum_{l \in \mathcal{N}_{DC}(k)} g_{kl} V_{k\_DC} V_{l\_DC}, \forall k \in \mathcal{N}_{DC} \\
&2s_k = \gamma_k q_{kl}^2 + \beta_k \|q_{kl}\| + \delta_k, \forall (k,l) \in \mathcal{D} \\
&P_{lm\_DC} = g_{lm}(V_{l\_DC}^2 - V_{l\_DC} V_{m\_DC}), \forall (l,m) \in \mathcal{I}_{DC} \\
&V_{k\_DC}^{\min} \leq V_{k\_DC} \leq V_{k\_DC}^{\max}, \forall k \in \mathcal{N}_{DC} \\
&P_{k\_DC}^{\min} \leq P_{k\_DC} \leq P_{k\_DC}^{\max}, \forall k \in \mathcal{N}_{DC} \\
&|P_{lm\_DC}| \leq P_{lm\_DC}^{\max}, \forall (l,m) \in \mathcal{I}_{DC} \\
&|q_{kl}| \leq q_{kl}^{\max}, \forall (k,l) \in \mathcal{D}
\end{aligned} \quad (6)$$

where $a_{kl}$ is the flow sequence value for DC/DC converter connected with bus $k$ and $l$. $a_{kl}$ equals to 1 if the power flows from $k$ to $l$ in the DC/DC converter, otherwise, it would be -1. $g_{kl}$ is conductance for the line connected between DC bus $k$ and $l$. In this paper, the AC/DC converter connected with DC bus 1 in Figure 1 are considered as master converter. Active power injection upper limits for DC buses are related with the capacities of AC/DC converters, while the lower limits are concerned with the least power delivered from AC grid to DC grid, or vice versa. However, under some specific circumstances, the limits are fixed. For example, as for the bus 6 in Figure 1 connected with offshore wind farm, because the power generated by the wind costs less than traditional generators, all the power generated from wind energy should be dispatched first, then the DC power injection limits $P_{6\_DC}^{\min} = P_{6\_DC}^{\max} = \eta_6 P_{wind\_6}$, while for bus 7 and 8, because there's no DC load and there are no power conversion in these two nodes, then $P_{7\_DC}^{\min} = P_{7\_DC}^{\max} = P_{8\_DC}^{\min} = P_{8\_DC}^{\max} = 0$.

The DC grid problem shown in (6) is not convex either because the power balance equation and converter loss equation are not affine equality constraints. In addition, there are many different forms for the objective function of MT-HVDC system. In this paper, the goal is to minimize the losses in the HVDC transmission lines with shown constraints in equation (6).

## IV. CONVEX RELAXATION OF OPTIMAL POWER FLOW PROBLEM BASED ON SEMIDEFINITE PROGRAMMING

Semidefinite programming is one of the optimization technique in convex optimization field, where the optimization variables are symmetric positive semidefinite matrices. In addition, the inequality constraints for SDP problem are the combination of linear matrix inequalities (LMIs) [13]. What's more, it is capable to solve problems with hundreds or thousands of variables in few seconds by applying interior point method [13]. Therefore, it is a very promising approach for power systems applications where the real-time control is required.

### A. SDP Relaxation

Due to the aforementioned OPF problems are not convex, similar like paper [7], the convex relaxation based SDP is applied first, and some new admittance matrices employed for bus injection and voltage magnitude are introduced as follows.

$$\begin{aligned}
&Y_{k\_AC} = e_k e_k^T Y_{AC} \\
&Y_{lm\_AC} = (\bar{y}_{lm\_AC} + y_{lm\_AC}) e_l e_l^T - y_{lm\_AC} e_l e_m^T \\
&\mathbf{Y_{k\_AC}} = \frac{1}{2}\begin{bmatrix} \operatorname{Re}\{Y_{k\_AC} + Y_{k\_AC}^T\} & \operatorname{Im}\{Y_{k\_AC}^T - Y_{k\_AC}\} \\ \operatorname{Im}\{Y_{k\_AC} - Y_{k\_AC}^T\} & \operatorname{Re}\{Y_{k\_AC} + Y_{k\_AC}^T\} \end{bmatrix} \\
&\mathbf{Y_{lm\_AC}} = \frac{1}{2}\begin{bmatrix} \operatorname{Re}\{Y_{lm\_AC} + Y_{lm\_AC}^T\} & \operatorname{Im}\{Y_{lm\_AC}^T - Y_{lm\_AC}\} \\ \operatorname{Im}\{Y_{lm\_AC} - Y_{lm\_AC}^T\} & \operatorname{Re}\{Y_{lm\_AC} + Y_{lm\_AC}^T\} \end{bmatrix} \quad (7a)\\
&\bar{\mathbf{Y}}_{\mathbf{k\_AC}} = \frac{-1}{2}\begin{bmatrix} \operatorname{Im}\{Y_{k\_AC} + Y_{k\_AC}^T\} & \operatorname{Re}\{Y_{k\_AC} - Y_{k\_AC}^T\} \\ \operatorname{Re}\{Y_{k\_AC}^T - Y_{k\_AC}\} & \operatorname{Im}\{Y_{k\_AC} + Y_{k\_AC}^T\} \end{bmatrix} \\
&\mathbf{M_{k\_AC}} = \begin{bmatrix} e_k e_k^T & 0 \\ 0 & e_k e_k^T \end{bmatrix}
\end{aligned}$$

$$\overline{\mathbf{Y}}_{\mathbf{lm\_AC}} = \frac{-1}{2}\begin{bmatrix} \text{Im}\{Y_{lm\_AC} + Y_{lm\_AC}^T\} & \text{Re}\{Y_{lm\_AC} - Y_{lm\_AC}^T\} \\ \text{Re}\{Y_{lm\_AC}^T - Y_{lm\_AC}\} & \text{Im}\{Y_{lm\_AC} + Y_{lm\_AC}^T\} \end{bmatrix}$$

$$\mathbf{X}_{\mathbf{AC}} = \begin{bmatrix} \text{Re}\{\mathbf{V}_{\mathbf{AC}}\}^T & \text{Im}\{\mathbf{V}_{\mathbf{AC}}\}^T \end{bmatrix}^T$$

$$\mathbf{W}_{\mathbf{AC}} = \mathbf{X}_{\mathbf{AC}}\mathbf{X}_{\mathbf{AC}}^{\mathbf{T}}$$

$$P_{k\_AC}^{\min} = P_{Gk\_AC}^{\min} - P_{Dk}, \forall k \in \mathscr{G}_{AC}$$

$$P_{k\_AC}^{\max} = P_{Gk\_AC}^{\max} - P_{Dk}, \forall k \in \mathscr{G}_{AC} \quad (7b)$$

$$Q_{k\_AC}^{\min} = Q_{Gk\_AC}^{\min} - Q_{Dk}, \forall k \in \mathscr{G}_{AC}$$

$$Q_{k\_AC}^{\max} = Q_{Gk\_AC}^{\max} - Q_{Dk}, \forall k \in \mathscr{G}_{AC}$$

$$P_{k\_AC}^{\min} = P_{k\_AC}^{\max} = -P_{Dk}, \forall k \in \mathscr{N}_{AC} \setminus \mathscr{G}_{AC}$$

$$Q_{k\_AC}^{\min} = Q_{k\_AC}^{\max} = -Q_{Dk}, \forall k \in \mathscr{N}_{AC} \setminus \mathscr{G}_{AC}$$

According to [7], the classical AC OPF problem in (1) could be transformed as follows:

$$\underset{\mathbf{W}_{\mathbf{AC}}}{minimize} \quad \sum_{k \in \mathscr{G}_{AC}} \alpha_k + P_{loss\_AC} \quad (8a)$$

*subject to*

$$P_{k\_AC}^{\min} \le \text{Tr}\{\mathbf{Y}_{\mathbf{k\_AC}}\mathbf{W}_{\mathbf{AC}}\} \le P_{k\_AC}^{\max}, \forall k \in \mathscr{N}_{AC} \quad (8b)$$

$$Q_{k\_AC}^{\min} \le \text{Tr}\{\overline{\mathbf{Y}}_{\mathbf{k\_AC}}\mathbf{W}_{\mathbf{AC}}\} \le Q_{k\_AC}^{\max}, \forall k \in \mathscr{N}_{AC} \quad (8c)$$

$$(V_{k\_AC}^{\min})^2 \le \text{Tr}\{\mathbf{M}_{\mathbf{k\_AC}}\mathbf{W}_{\mathbf{AC}}\} \le (V_{k\_AC}^{\max})^2, \forall k \in \mathscr{N}_{AC} \quad (8d)$$

$$(V_{k\_AC}^{\min})^2 \le \text{Tr}\{\mathbf{M}_{\mathbf{k\_AC}}\mathbf{W}_{\mathbf{AC}}\} \le (V_{k\_AC}^{\max})^2, \forall k \in \mathscr{N}_{AC} \quad (8e)$$

$$\begin{bmatrix} -S_{lm\_AC\max}^2 & \text{Tr}\{\mathbf{Y}_{\mathbf{lm\_AC}}\mathbf{W}_{\mathbf{AC}}\} & \text{Tr}\{\overline{\mathbf{Y}}_{\mathbf{lm\_AC}}\mathbf{W}_{\mathbf{AC}}\} \\ \text{Tr}\{\mathbf{Y}_{\mathbf{lm\_AC}}\mathbf{W}_{\mathbf{AC}}\} & -1 & 0 \\ \text{Tr}\{\overline{\mathbf{Y}}_{\mathbf{lm\_AC}}\mathbf{W}_{\mathbf{AC}}\} & 0 & -1 \end{bmatrix} \preceq 0, \forall(l,m) \in \mathscr{Z}_{AC} \quad (8f)$$

$$\begin{bmatrix} -S_{lm\_AC\_\max}^2 & \text{Tr}\{\mathbf{Y}_{\mathbf{ml\_AC}}\mathbf{W}_{\mathbf{AC}}\} & \text{Tr}\{\overline{\mathbf{Y}}_{\mathbf{ml\_AC}}\mathbf{W}_{\mathbf{AC}}\} \\ \text{Tr}\{\mathbf{Y}_{\mathbf{ml\_AC}}\mathbf{W}_{\mathbf{AC}}\} & -1 & 0 \\ \text{Tr}\{\overline{\mathbf{Y}}_{\mathbf{ml\_AC}}\mathbf{W}_{\mathbf{AC}}\} & 0 & -1 \end{bmatrix} \preceq 0, \forall(m,l) \in \mathscr{Z}_{AC} \quad (8g)$$

$$\begin{bmatrix} c_{k1}\text{Tr}\{\mathbf{Y}_{\mathbf{k\_AC}}\mathbf{W}_{\mathbf{AC}}\} - \alpha_k + c_{k0} + c_{k1}P_{D_{k\_AC}} & \sqrt{c_{k2}}\text{Tr}\{\mathbf{Y}_{\mathbf{k\_AC}}\mathbf{W}_{\mathbf{AC}}\} + \sqrt{c_{k2}}P_{D_{k\_AC}} \\ \sqrt{c_{k2}}\text{Tr}\{\mathbf{Y}_{\mathbf{k\_AC}}\mathbf{W}_{\mathbf{AC}}\} + \sqrt{c_{k2}}P_{D_{k\_AC}} & -1 \end{bmatrix} \preceq 0 \quad (8h)$$

$$, \forall k \in \mathscr{G}_{AC}$$

$$P_{loss\_AC} = \sum_{(l,m) \in \mathscr{Z}_{AC}}^{n_{ACLines}} (\text{Tr}\{\mathbf{Y}_{\mathbf{lm\_AC}}\mathbf{W}_{\mathbf{AC}}\} + \text{Tr}\{\mathbf{Y}_{\mathbf{ml\_AC}}\mathbf{W}_{\mathbf{AC}}\}) \quad (8i)$$

$$\mathbf{W}_{\mathbf{AC}} \succeq 0 \quad (8j)$$

$$\mathbf{W}_{\mathbf{AC}}(\text{ref}+n_{AC}, \text{ref}+n_{AC}) = 0 \quad (8k)$$

$$\text{rank}(\mathbf{W}_{\mathbf{AC}}) = 1 \quad (8l)$$

The equation (8k) is the constraint for the slack bus, where the imaginary part of the slack bus voltage should be 0. Because the constraint of equation (8l), this problem is still not convex, however, if this equation is removed, then the problem could be solved by SDP. Also, if this SDP relaxation is feasible and the rank of $\mathbf{W}_{\mathbf{AC}}$ is coincidently 1, then the problem is solved, and furthermore, this solution could be seen as globally optimal point. From paper [7]-[9], it is proved that the feasible solution of AC OPF problem from the SDP relaxation could be the recovered if the rank of $\mathbf{W}_{\mathbf{AC}}$ is less than 2.

Similarly, the OPF problem for MT-HVDC system could be formulated. The conductance matrices are defined as follows:

$$Y_{i\_DC} = e_i e_i^T Y_{DC}$$

$$Y_{ft\_DC} = g_{ft} e_f e_f^T - g_{ft} e_f e_t^T$$

$$\mathbf{Y}_{\mathbf{i\_DC}} = \frac{1}{2}(Y_{i\_DC} + Y_{i\_DC}^T)$$

$$\mathbf{Y}_{\mathbf{ft\_DC}} = \frac{1}{2}(Y_{ft\_DC} + Y_{ft\_DC}^T)$$

$$\overline{\mathbf{Y}}_{\mathbf{i\_DC}} = 0 \quad (9)$$

$$\overline{\mathbf{Y}}_{\mathbf{ft\_DC}} = 0$$

$$\mathbf{M}_{\mathbf{i\_DC}} = e_i e_i^T$$

$$\mathbf{X}_{\mathbf{DC}} = \mathbf{V}_{\mathbf{DC}}$$

$$\mathbf{W}_{\mathbf{DC}} = \mathbf{X}_{\mathbf{DC}}\mathbf{X}_{\mathbf{DC}}^{\mathbf{T}}$$

Then the OPF problem for MT-HVDC system in equation (6) is shown as follows:

$$\underset{\mathbf{W}_{\mathbf{DC}}, S, Q}{minimize} \quad P_{loss\_DC} \quad (10a)$$

*subject to*

$$P_{loss\_DC} = \sum_{(l,m) \in \mathscr{Z}_{DC}}^{n_{DClines}} (\text{Tr}\{\mathbf{Y}_{\mathbf{ft\_DC}}\mathbf{W}_{\mathbf{DC}}\} + \text{Tr}\{\mathbf{Y}_{\mathbf{tf\_DC}}\mathbf{W}_{\mathbf{DC}}\}) + \sum_{i \in \mathscr{N}_{DC}}^{n_{AC\_Conv}} S_i \quad (10b)$$

$$P_{k\_DC}^{\min} \le \text{Tr}\{\mathbf{Y}_{\mathbf{k\_DC}}\mathbf{W}_{\mathbf{DC}}\} \le P_{k\_DC}^{\max}, \forall k \in \mathscr{N}_{DC} \quad (10c)$$

$$(V_{k\_DC}^{\min})^2 \le \text{Tr}\{\mathbf{M}_{\mathbf{k\_DC}}\mathbf{W}_{\mathbf{DC}}\} \le (V_{k\_DC}^{\max})^2, \forall k \in \mathscr{N}_{DC} \quad (10d)$$

$$|\text{Tr}\{\mathbf{Y}_{\mathbf{ft\_DC}}\mathbf{W}_{\mathbf{DC}}\}| \le P_{lm\_DC\max}, \forall k \in \mathscr{N}_{DC} \quad (10e)$$

$$|\text{Tr}\{\mathbf{Y}_{\mathbf{tf\_DC}}\mathbf{W}_{\mathbf{DC}}\}| \le P_{ml\_DC\max}, \forall k \in \mathscr{N}_{DC} \quad (10f)$$

$$P_{k\_DC} = S_k - A_{d\_k} \times Q_k - \text{Tr}\{\mathbf{Y}_{\mathbf{k\_DC}}\mathbf{W}_{\mathbf{DC}}\} \quad (10g)$$

$$2S \ge |A_d| \times (\gamma \cdot (Q \circ Q) + \beta \cdot |Q|) \quad (10h)$$

$$|Q| \le Q_{\max} \quad (10i)$$

$$\mathbf{W}_{\mathbf{DC}} \succeq 0 \quad (10j)$$

$$\text{rank}(\mathbf{W}_{\mathbf{DC}}) = 1 \quad (10k)$$

The OPF problem in MT-HVDC system is convex if the equation (10k) is removed. If the rank of $\mathbf{W}_{\mathbf{DC}}$ in the corresponding feasible solution is 1, then the global optimum solution is achieved. In this paper, for the hybrid AC and MT-HVDC system with offshore wind farm integration, the objective function is to minimize the cost of generators and losses in the system. Then the objective function is shown in equation (11). What's more, the equation (3) for the converter balance constraint and the equation (4) power converter capacity constraint in SDP format are shown in (12) and (13). Then the problem for the hybrid AC MT-HVDC system as SDP relaxation format is formulated as equation (11) with the constraints (8b)-(8k), (10b)-(10j), (12), (13).

$$\underset{\mathbf{W_{AC}},\mathbf{W_{DC}},S,Q}{minimize} \quad \sum_{k\in \mathcal{I}_{AC}} \alpha_k + P_{loss\_AC} + P_{loss\_DC} \quad (11)$$

$$\eta_k \times \text{Tr}\{\mathbf{Y_{k\_AC}W_{AC}}\} = -\text{Tr}\{\mathbf{Y_{k\_DC}W_{DC}}\}, \forall k \in \mathcal{O} \quad (12)$$

$$\begin{bmatrix} -S_{k\_conv}^2 & \text{Tr}\{\mathbf{Y_{k\_AC}W_{AC}}\} & \text{Tr}\{\mathbf{\overline{Y}_{k\_AC}W_{AC}}\} \\ \text{Tr}\{\mathbf{Y_{k\_AC}W_{AC}}\} & -1 & 0 \\ \text{Tr}\{\mathbf{\overline{Y}_{k\_AC}W_{AC}}\} & 0 & -1 \end{bmatrix} \preceq 0, \forall k \in \mathcal{O} \quad (13)$$

### B. Voltage Recovery

From [7], it is proved that if the rank of matrix $\mathbf{W_{AC}}$ is one, then there's zero duality gap for AC OPF problem by applying the SDP relaxation, and if the rank is 2, then it's possible to recover the original solution and the global optimum solution is achieved. If the rank is bigger than 2, then the original solution couldn't be solved. From the paper [14], if the eigenvalue ratio between the largest eigenvalue and the third largest eigenvalue is bigger than $10^5$, then then rank of matrix $\mathbf{W_{AC}}$ could be seen less than or equal to 2. After achieving the feasible matrices $\mathbf{W_{AC}}$ and $\mathbf{W_{DC}}$ with suitable ranks, the voltages could be recovered. The main approach to recover the optimal voltage vector is to use the eigenvalues and corresponding eigenvectors. If the rank of matrices $\mathbf{W_{AC}}$ or $\mathbf{W_{DC}}$ is 1, then the optimal voltage vector is shown as follows:

$$\begin{aligned} X_{opt\_AC} &= \sqrt{\lambda_{1\_AC}} E_{1\_AC} \\ V_{opt\_AC} &= X_{opt\_AC}(1:n_{AC}) + jX_{opt\_AC}(n_{AC}+1:2n_{AC}) \\ V_{opt\_DC} &= \sqrt{\lambda_{1\_DC}} E_{1\_DC} \end{aligned} \quad (14)$$

If the rank of $\mathbf{W_{AC}}$ is 2, then the recovery equation is [8]:

$$\begin{aligned} X_{opt\_AC} &= \sqrt{\lambda_{1\_AC}} E_{1\_AC} + \sqrt{\lambda_{2\_AC}} E_{2\_AC} \\ V_{opt\_AC} &= X_{opt\_AC}(1:n_{AC}) + jX_{opt\_AC}(n_{AC}+1:2n_{AC}) \end{aligned} \quad (15)$$

Both the recovered voltage vectors for AC system are in complex format for equation (14) and (15).

## V. CASE STUDY

In this paper, the MT-HVDC system with integrated offshore wind farm for the simulation is the same system shown in Figure 1, while the AC system is the combination of IEEE 39 bus system and IEEE 9 bus system, where there's one line connected bus 6 in IEEE 39 system with bus 7 in IEEE 9 system, and it is shown in Figure 4. Also, the apparent power limits for this transmission line is 30MVA.

The bus number for IEEE 9 bus system is re-numbered after IEEE 39 bus system. The objective function is to minimize the costs of the generators and the losses in the system shown in equation (11) with the constraints shown in (8b)-(8k), (10b)-(10j), (12), (13). The AC bus numbers connected with MT-HVDC system are shown in Figure 1 with regrouping numbers. The output of the offshore wind farm is set as 700MW and the parameters of converters are shown in table I. The HVDC line parameters comes from [6]. And the controlled voltage for AC/DC converter is set as 0.98 p.u.

TABLE I  CONVERTER PARAMETERS OF THE MT-HVDC SYSTEM

| Bus | Type | Capacity (MVA) | Rated Voltage (kV) | $P^{min}$ | $P^{max}$ |
|---|---|---|---|---|---|
| 1 | AC/DC | 500 | 345 | -300 | 300 |
| 2 | AC/DC | 360.55 | 345 | -300 | 300 |
| 3 | AC/DC | 224 | 345 | -300 | 300 |
| 4 | AC/DC | 283 | 345 | -200 | 200 |
| 5 | AC/DC | 283 | 345 | -200 | 200 |
| 6 | AC/DC | 1000 | 345 | 700 | 700 |
| 7 | - | 0 | 345 | 0 | 0 |
| 8 | - | 0 | 345 | 0 | 0 |
| 3-8 | DC/DC | 200 | 345 | -200 | 200 |

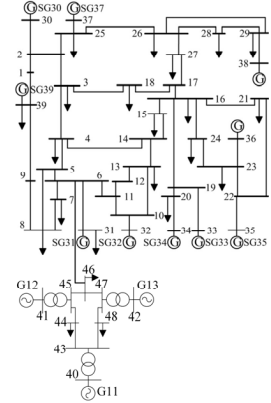

Figure 4. AC system for case study

The hybrid system with MT-HVDC case is simulated in MATLAB, where the YALMIP [15] is used for the model establishment. In addition, the system data is adopted from matpower6.0 [16]. MOSEK is used as the solver for the problem. The function runsdpopf() provided by [8] in matpower6.0 is run first for AC system shown in Figure 4 as a comparing result with the hybrid AC and MT-HVDC system. The eigenvalue ratio of matrix $\mathbf{W_{DC}}$ is $2.2*10^7$, which approximately guarantee the rank of $\mathbf{W_{DC}}$ is 1. In this paper, $\delta_1=0$, $\beta_1=0.05$, $\gamma_1=0.03$. From the simulation, the cost function results for AC system is \$46924.85/hr, and the total loss is 49.3 MW. Because of the zero cost for 700MW wind farm integration in this paper for the hybrid system, the cost is reduced to \$36949.28/hr. The comparison results between AC system and the hybrid AC MT-HVDC system in the aspect of costs and losses are shown in table II. And the other simulation results are shown from Figure 5 to 10. As for the power flow in MT-HVDC system, the test results and the line flow limits used for the simulation are shown in table III.

TABLE II COMPARISON RESULTS FOR AC SYSTEM AND HYBRID AC MT-HVDC SYSTEM

|  | AC System | Hybrid System |
|---|---|---|
| Cost (\$/h) | 46924.85 | 36949.28 |
| AC system Loss (MW) | 49.3 | 37.52 |

TABLE III POWER FLOW IN THE MT-HVDC SYSTEM

| From bus | To bus | Power Flow (MW) | Flow Limit (MW) |
|---|---|---|---|
| 2 | 6 | -298.634 | 300 |
| 6 | 2 | 299.903 | 300 |
| 1 | 6 | -174.8027 | 300 |
| 6 | 1 | 176.3107 | 300 |
| 6 | 3 | 223.7834 | 300 |
| 3 | 6 | -222.8413 | 300 |
| 1 | 3 | -215.1381 | 300 |
| 3 | 1 | 216.0205 | 300 |
| 1 | 4 | 89.9941 | 90 |
| 4 | 1 | -89.908 | 90 |
| 4 | 5 | 26.3661 | 300 |
| 5 | 4 | -26.4035 | 300 |
| 5 | 7 | -42.2165 | 300 |
| 7 | 5 | 42.2285 | 300 |
| 8 | 7 | 42.2094 | 300 |
| 7 | 8 | -42.2288 | 300 |

| 3 | 8 | 2.7553 | 200 |

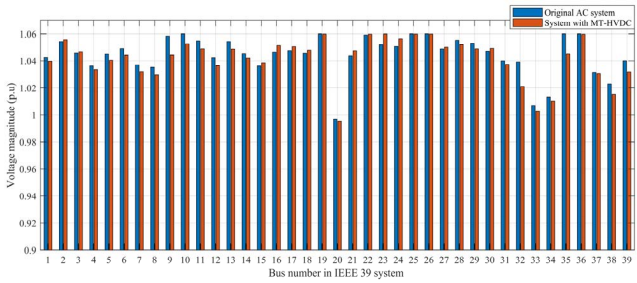

Figure 5. Comparision results voltage magnitude in p.u. for IEEE 39 bus

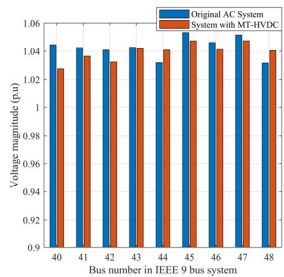

Figure 6. Comparision results voltage magnitude in p.u. for IEEE 9 bus

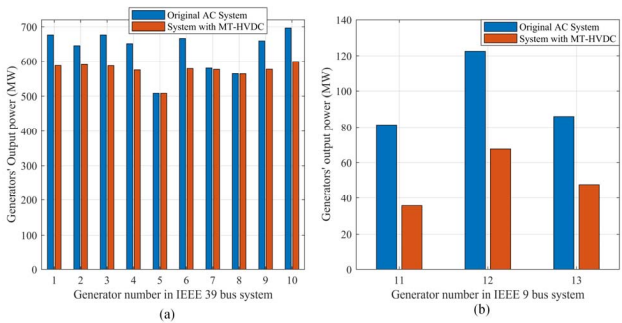

Figure 7. Comparision results for generator outputs. (a) IEEE 39 (b) IEEE 9

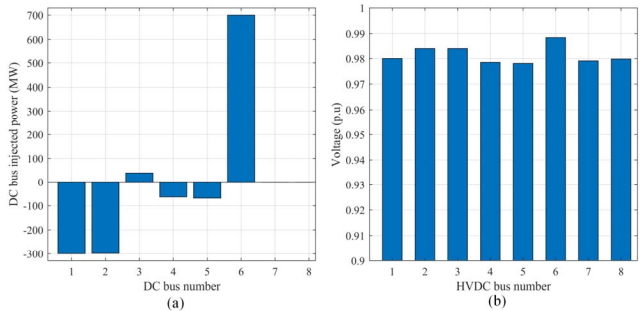

Figure 8. Comparision results for generator outputs. (a) IEEE 39 (b) IEEE 9

From table II and Figure 7 and 8, because of the offshore wind farm integration with the system, the generators outputs are reduced, which leads to less costs for the operation. Furthermore, only the DC bus 6 and 3 are positive power injection in the MT-HVDC system, while other AC/DC converters deliver the offshore wind farm's output to the AC system.

VI. CONCLUSION

In this paper, the convex relaxation for MT-HVDC system with integrated offshore wind farm is presented by applying the semidefinite programming model. The combined IEEE 39 bus and IEEE 9 bus system is used as the AC system for the simulation. In addition, the DC/DC converter model is considered in MT-HVDC system, which is an important component to control the power flow in HVDC transmission lines. The simulation results validate the effectiveness of the convex relaxation model.